\documentclass[final,authoryear,12pt]{elsarticle}
\pdfoutput=1

\usepackage{color}
\usepackage{booktabs}
\usepackage{dcolumn}
\usepackage{array}
\usepackage{mathtools}
\usepackage{graphicx}
\usepackage{amssymb}
\usepackage{amsmath}
\usepackage{setspace}

\begin{document}

\setlength{\extrarowheight}{10pt}

\begin{frontmatter}

\title{DeepMRSeg: A convolutional deep neural network for anatomy and abnormality segmentation on MR images}

\author{Jimit Doshi\corref{cor1}\fnref{label1}}
\author{Guray Erus\fnref{label1}}
\author{Mohamad Habes\fnref{label1}}
\author{Christos Davatzikos\fnref{label1}}
\address[label1]{Center for Biomedical Image Computing and Analytics, University of Pennsylvania, Philadelphia, PA, USA}
\cortext[cor1]{Corresponding author. Address: University of Pennsylvania, Richards Building, 3700 Hamilton Walk, 7th Floor, Philadelphia, PA 19104. Tel: +1 215 662 7362. E-mail: jimit.doshi@pennmedicine.upenn.edu}

\pagebreak

\begin{abstract}

Segmentation has been a major task in neuroimaging. A large number of automated methods have been developed for segmenting healthy and diseased brain tissues. In recent years, deep learning techniques have attracted a lot of attention as a result of their high accuracy in different segmentation problems. We present a new deep learning based segmentation method, DeepMRSeg, that can be applied in a generic way to a variety of segmentation tasks. The proposed architecture combines recent advances in the field of biomedical image segmentation and computer vision. We use a modified UNet architecture that takes advantage of multiple convolution filter sizes to achieve multi-scale feature extraction adaptive to the desired segmentation task. Importantly, our method operates on minimally processed raw MRI scan. We validated our method on a wide range of segmentation tasks, including white matter lesion segmentation, segmentation of deep brain structures and hippocampus segmentation. We provide code and pre-trained models to allow researchers apply our method on their own datasets.

\end{abstract}


\end{frontmatter}

% \clearpage

\section{Introduction}
\label{label_Introduction}

Segmentation has been a major task in medical image analysis since the early years of the field, as it enables quantification of normal and abnormal anatomical regions both for individual assessments and for comparative group analyses \citep{GONZALEZVILLA201645}. In neuroimaging, multiple automated methods have been developed for various problems, such as brain extraction, segmentation of anatomical regions of interest (ROIs), white matter lesion (WML) segmentation and segmentation of brain tumor sub-regions. Importantly, each of these problems have their own specific challenges, mainly due to variations in image modalities and imaging signatures that best characterize target regions. These variations motivated development of a large number of distinct task-specific segmentation methods \citep{Kalavathi16,ANBEEK20041037,iglesias14,GORDILLO20131426,despotovic15}.

Machine learning has played a key role in enabling novel methods that achieved accuracy comparable  to, or surpassing human raters. In the commonly used supervised learning framework, examples with ground-truth labels are presented to the learning algorithm in order to construct a model that learns imaging patterns that characterize the target segmentations. The model is then applied to new scans for segmenting the target areas on them. Common supervised learning techniques, such as the support vector machines, have obtained very promising results. However, they require a number of sophisticated preprocessing and feature elimination/selection steps, making them vulnerable to negative effects of scanner variations and limiting their widespread usage in clinical settings.

In recent years, deep learning techniques have attracted a lot of attention as a result of their state-of-the-art performance in multiple major problems in computer vision and image analysis \citep{Guo2018}. In neuroimaging, convolutional neural networks (CNN) have started to gain popularity, with successful applications on various image recognition tasks \citep{KamnitsasLNSKMR16, Akkus17, Anwar18}. CNNs are deep neural networks designed to take advantage of the 2D or 3D structure of the input images. An input image passes through a series of convolution layers followed by pooling layers, which are acting together as filters to extract multiple translation invariant local features, without need for the manual feature engineering traditionally required.

In this paper we present DeepMRSeg, a deep learning based segmentation method that can be applied in a generic way to a variety of segmentation tasks. Our method uses a modified version of the UNet architecture \citep{RonnebergerFB15}, with ResNet \citep{HeZRS15} and modified Inception-ResNet-A \citep{SzegedyIV16} blocks in the encoding and decoding paths, taking advantage of recent advances in biomedical image segmentation and image classification. The residual connections allow the UNet architecture to learn residual mappings, while multiple branches of convolutional layers with different kernel sizes allow the network to learn multi-scale features that are adaptive to the segmentation task at hand. Also, we replace the maxpool operations in the UNet architecture with 1x1 convolutional filters to prevent the loss of fine boundary details that are important for a segmentation task. Importantly, our method operates on minimally processed raw MRI scans and it can be directly applied for different segmentation problems after training a model with the specific training data type. We validated our method on a wide range of segmentation tasks, including WML segmentation, segmentation of deep brain structures, and hippocampus segmentation. The DeepMRSeg method and trained models are provided on our online platform, IPP (https://ipp.cbica.upenn.edu/), to allow users to apply our methods and models without need for installing any software packages.

\section{Related Work}
\label{sec:RelWork}

UNet architecture was introduced by \cite{RonnebergerFB15}. UNet has been an important advancement in the application of deep CNNs to the problem of biomedical image segmentation. CNNs have been initially used on classification problems by mapping input images to output class labels. However, in segmentation tasks, the desired output is an image, e.g. a binary segmentation map. UNet extended the CNN architecture by supplementing the usual contracting or encoding path with a symmetric expanding or decoding path, where pooling operators are replaced by upsampling operators. The encoding path allows the architecture to learn spatially relevant context information and the decoding path adds the precise localization back to the architecture, leading to  a final segmentation image as output of the model.

A straightforward way for improving the performance of deep neural networks is to increase the network  size, either by increasing the depth (number of layers) or the width (number of units at each layer). The ResNet architecture \citep{HeZRS15}, was introduced to address an important limitation of deep learning, known as the degradation problem. As a network goes deeper, the gradient of the error function used for the weight updates may vanish, resulting in degrading accuracy. The main idea in ResNet is to learn a ``residual mapping'' instead of directly learning the desired underlying mapping to address this problem. For this purpose, shortcut connections that skip one or more layers are introduced. These shortcut connections are identity mappings that simply add their outputs to the outputs of the stacked convolutional layers, thus making the layer learn the residual. ResNet allowed training deeper networks, while providing significantly faster convergence.
Large convolution operations are computationally expensive, as the network will have a larger number of parameters to learn. The Inception framework \citep{SzegedyLJSRAEVR14}, was proposed to overcome this limitation by introducing sparsity in the network architecture. The main idea is to constrain the network to lower dimensions and group them into highly correlated filter units. The authors used different types of convolutional filter sizes (1x1, 3x3, 5x5) instead of a single filter size (3x3), allowing the network to learn different representations of the input image. This strategy helped to reduce the number of parameters/connections, thus reducing the complexity of the network and allowing the network to go wider while keeping the computational budget constant.

As noted by \cite{NIPS2012_4824}, when convolutional filters are arranged in different groups, the network can learn distinct features from each group, with low correlation between the learned features across groups. This was demonstrated in AlexNet, where the network consistently identified color-agnostic and color-specific features in different filter groups. The same concept also applies to Inception network, where through the use of different convolutional filters at a single layer the network  learns feature representations at different resolution levels.

\section{Network Architecture}

\subsection{Overview}

\begin{figure}[!htbp]
  \begin{center}
    \includegraphics[width=\textwidth]{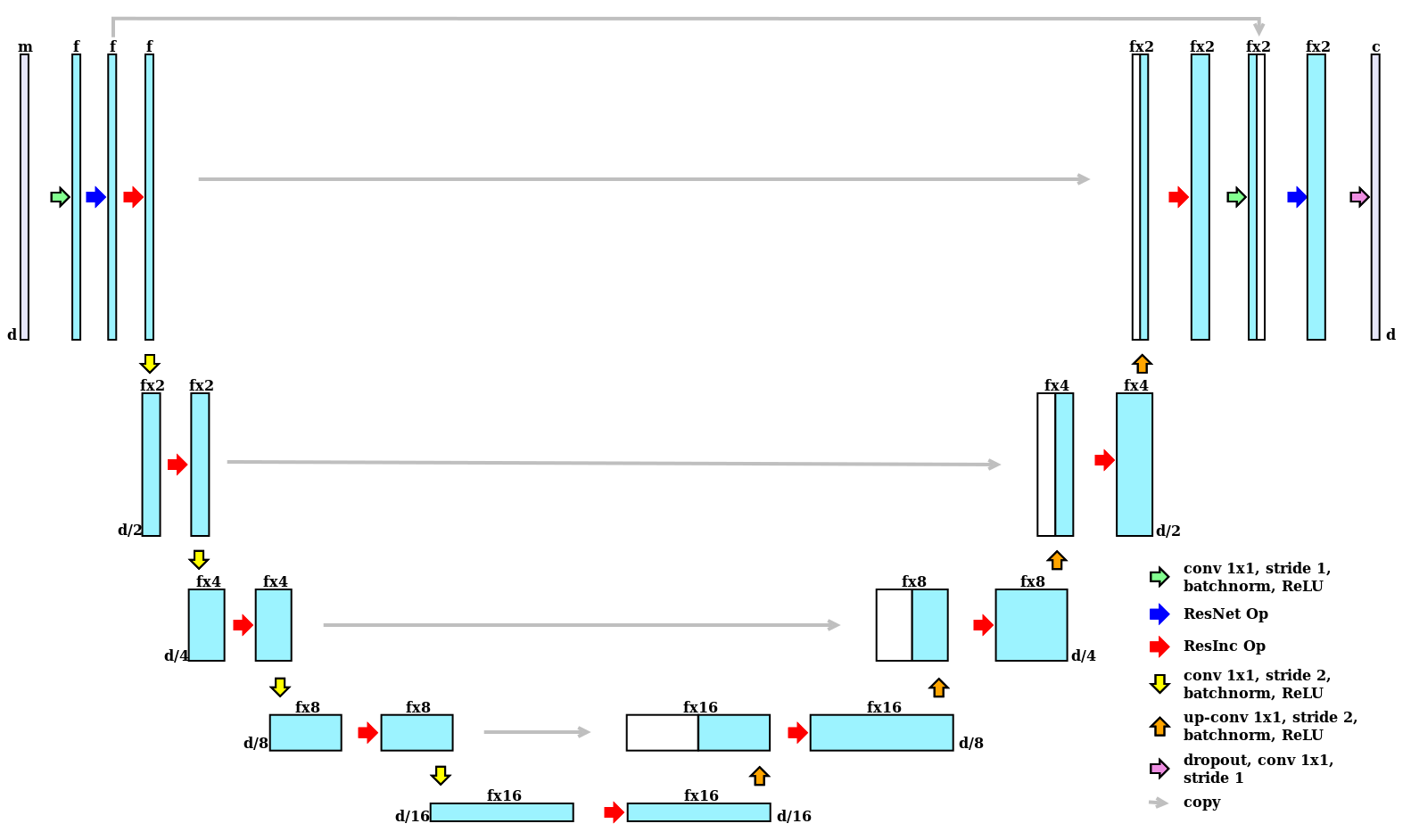}
    \caption{Overview of the DeepMRSeg network architecture}
    \label{Fig1_Network}
  \end{center}
\end{figure}

An overview of the proposed architecture is illustrated in figure \ref{Fig1_Network}. DeepMRSeg is built upon components that combine ideas from recent advances in the field, as described in the previous section. The network architecture consists of an encoding path and a corresponding decoding path, as in UNet \citep{RonnebergerFB15}, followed by a voxel-wise, multi-class soft-max classifier to produce class probabilities for each voxel independently. An initial projection layer transforms the input feature maps (m) into the desired number of features (f). These features go through a pre-encoding block, consisting of ResNet blocks that extract various features from the input images and forms the input to the UNet. The encoding path of the network consists of encoder blocks that operate at different feature map resolutions. At each layer, the feature maps are subsampled using the ``transition down'' operation and they are fed into a ResInc block. The size of the feature maps decreases at each layer, while their receptive field increases, thus encoding more context information into the network. The decoding pathway includes up-sampling operations symmetric to the encoding blocks, coupled with ResInc blocks. Individual components of the DeepMRSeg architecture are explained in more details below.

\subsection{Projection Layer}

This layer is used at the start of the network to project the input image channels, or modalities (m), to a set of feature maps (f). This is accomplished with a convolution layer with a kernel size of 1x1 and stride 1. Intuitively, this layer learns a linear combination of the input channels and projects it onto the desired number of feature maps required for subsequent layers. The convolution in this layer is followed by batch normalization \citep{IoffeS15} and ReLU activation \citep{Nair10}.

\subsection{ResNet Module}

\begin{figure}[!htbp]
  \begin{center}
    \includegraphics[width=13cm]{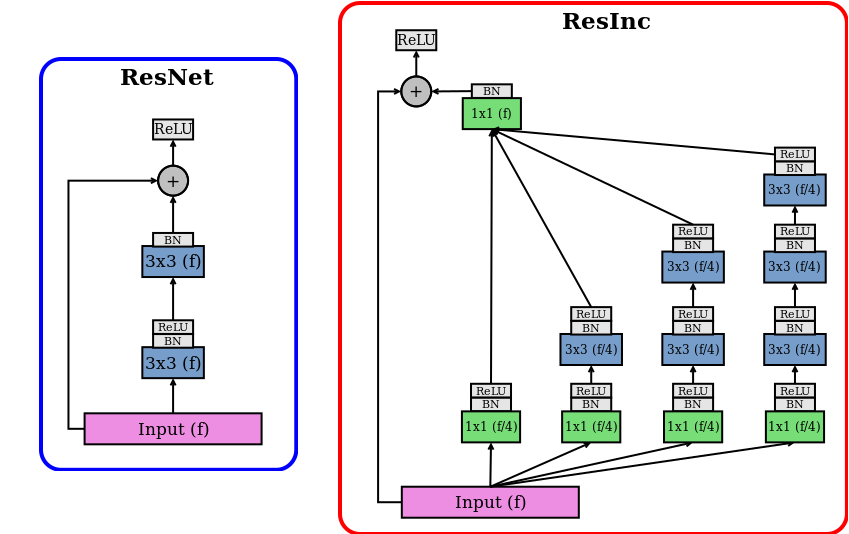}
    \caption{The ResNet and ResInc block architectures. The input data is the set of ``f'' feature maps obtained from the output of the previous layer.}
    \label{Fig2_ResNetResInc}
  \end{center}
\end{figure}

Following the general design principles described in \citep{SzegedyVISW15}, we avoid representational bottlenecks with extreme compression early in the network. To achieve this, we add the traditional ResNet block (fig. \ref{Fig2_ResNetResInc}.A) before the encoding pathway to ensure that the input features go through a few layers of convolutions before being fed into the ResInc block.

\subsection{ResInc Module}

The ResInc module (fig. \ref{Fig2_ResNetResInc}.B), modified from the Inception-ResNet-A module of the Inception-ResNet-v2 architecture, is used at every level of the UNet architecture, coupled with the ``Transition Down'' operation. This module splits the incoming input feature maps into 4 branches and an identity mapping that is added back to the final output. Each branch reduces the input dimensionality from ``f'' feature maps to ``f/4'' feature maps. This introduces a bottleneck by reducing the dimensionality of the incoming features and also reduces the number of learnable parameters in the network. Each branch subsequently transforms the input features with a varying number of 3x3 convolution layers. This ensures that each branch learns a different representation of the input features by learning shallow as well as deeper features, and allows the subsequent layer to abstract features from different scales simultaneously. This property can be extremely useful when dealing with segmentation tasks for more complex or heterogeneous structures.
 
The 4 branches are concatenated to form a single feature vector that then goes through a 1x1 convolution layer before the residual connection is added. This 1x1 convolution acts as a linear weighting of the features learned in each branch. Thus, if a certain representation is not useful, it can be assigned a lower weight.

This configuration of the ResInc module has less than one-third the number of learnable parameters compared to the traditional ResNet block. This allows us to increase the width and depth of the network while keeping the number of learnable parameters low.

\subsection{Transition Down Blocks}

In the traditional UNet architecture, a maxpool layer is used to reduce the dimensionality of the feature maps. This allows the subsequent convolutional layers to have a larger field of view and therefore, more contextual information. This operation is used to achieve translation invariance over small spatial shifts in the input image. Adding several such operations in the network achieves more translation invariance for robust classification, but it can also lead to a loss of spatial resolution (boundary detail) of the input feature maps. This lossy representation of the boundary details is not desirable for segmentation tasks where boundary delineation is important.

Motivated by the work done by \citep{springenberg2014striving}, we have replaced the maxpool layer with a 1x1 convolution layer with stride 2, with the intuition that we let the network learn the parameters required for achieving the downsampling of the feature maps. This adds a few more learnable parameters in the network, resulting in a larger model size compared to maxpooling. However, while maxpooling works on subsampling within each feature map, the proposed convolution operation allows to model inter-feature dependencies.

The choice of kernel size was based on minimizing the number of learnable parameters outside of blocks containing residual connections. This operation serves the two tasks of subsampling the feature maps and doubling the number of feature maps simultaneously. This is followed by batch normalization and ReLU activation.

\subsection{Transition Up Blocks}

For upsampling the feature maps, we use a transposed convolution layer with a kernel size of 1 and stride 2. Upsampling the feature maps allows the addition of the skip connection from the previous layer from the encoding path. These skip connections are essential as they add the spatial localization information back into the network. Along with upsampling, the number of output feature maps are reduced to one-forth.

\subsection{Training Methodology}

The cost function to be minimized was a combination of softmax cross-entropy ($l_{CE}$), mean squared error ($l_{MSE}$) and soft Intersection Over Union ($l_{IOU}$). $l_{MSE}$ and $l_{IOU}$ were calculated between the one-hot encoded labels and the predicted soft max probabilities. The final loss, $l_{TOT}$ was an equal weighted sum of $l_{CE}$, $l_{MSE}$ and $l_{IOU}$. Adam optimizer was used to minimize the loss function ($l_{TOT}$). We started with a learning rate of $0.05$ and used a decayed learning rate schedule with a decay factor of $0.98$ every epoch. Data augmentation was done using randomized left/right flipping, followed by random translation, rotation and brightness/contrast adjustment.

\subsection{Evaluation Datasets}

We performed validation experiments on 3 different segmentation problems, specifically WML, mid-brain and hippocampus segmentation. We used publicly available datasets with ground-truth labels for each problem.  

\begin{figure}[!htbp]
  \begin{center}
    \includegraphics[width=11cm]{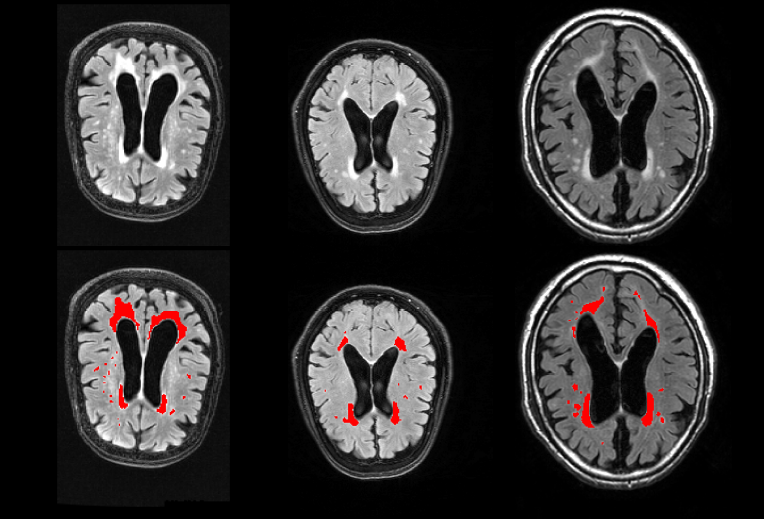}
    \caption{Example ground-truth segmentation for the WML.}
    \label{Fig_WML_Example}
  \end{center}
\end{figure}

\noindent \textbf{WML segmentation:} We used the training dataset that was provided as part of the MICCAI 2017 WML Segmentation Challenge \citep{Kuijf19}. This dataset included 3D T1-weighted and 2D multi-slice FLAIR scans from 60 subjects and manually delineated WML masks for these scans. The MRI scans were acquired from three different institutes/scanners: the University Medical Center (UMC), Utrecht, VU University Medical Centre (VU), Amsterdam, and the National University Health System (NUHS), Singapore. Manual segmentations were generated by one expert rater, following the STandards for ReportIng Vascular changes on nEuroimaging (STRIVE) protocol \citep{Wardlaw13} (Figure \ref{Fig_WML_Example}).

\begin{figure}[!htbp]
  \begin{center}
    \includegraphics[width=11cm]{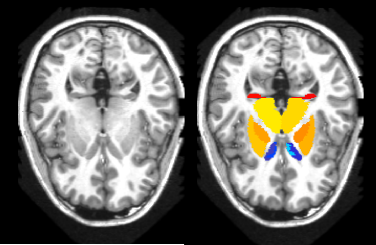}
    \caption{Example ground-truth segmentation for deep brain structures.}
    \label{Fig_Midbrain_Example}
  \end{center}
\end{figure}

\noindent \textbf{Deep brain segmentation:} We applied DeepMRSeg for segmentation of deep brain structures using the publicly available datset from MICCAI 2013 segmentation challenge \citep{asman2013miccai}. This dataset included T1-weighted scans of 35 subjects from OASIS project and corresponding manually created reference labels for 14 deep brain structures. The target regions of interest included accumbens, amygdala, caudate, hippocampus, pallidum, putamen and thalamus, separately for the left and the right hemispheres (fig \ref{Fig_Midbrain_Example}).

\begin{figure}[!htbp]
  \begin{center}
    \includegraphics[width=11cm]{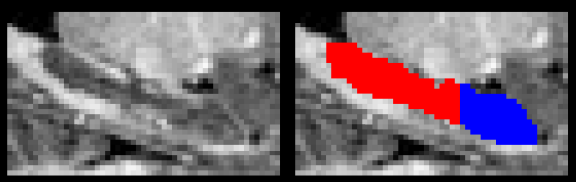}
    \caption{Example ground-truth segmentation for the hippocampus.}
    \label{Fig_Hippo1_Example}
  \end{center}
\end{figure}

\noindent \textbf{Hippocampus segmentation:} We also applied DeepMRSeg for segmenting the hippocampus, a structure critical in learning and memory, and particularly vulnerable to damage in early stages of AD \citep{Mu2011AdultHN}. We used the dataset provided as part of the Decathlon challenge, consisting of 3D T1-weighted MPRAGE scans of 195 subjects and manual hippocampus segmentations into two sub-regions, hippocampus tail and body \citep{simpson19} (Figure \ref{Fig_Hippo1_Example}).

\section{Evaluation Experiments and Metrics}

We compared DeepMRSeg against a modified UNet architecture where a batch normalization layer was added between the convolution and ReLU layers. This was used as the benchmark method to compare the proposed architecture against. The loss function, optimizer and data augmentation was the same as the one used for the proposed architecture. The two network models were trained using the appropriate type of labeled data for each specific segmentation task. We performed four-fold cross-validation in all experiments, with 25\% of the data left out for testing and the remaining data used for training and validation. This was repeated 20 times with randomization, giving us a robust estimate of the performance of the networks.

We quantified the performance of the networks using three complementary metrics that are commonly used for measuring the overlap between binary segmentation masks. We report the $F_1$ score (also known as the Dice coefficient), the $F_2$ score and the balanced accuracy between automated and expert delineated ground-truth segmentations. We calculated the scores individually for each subject and target ROI, and we reported mean and standard deviation of the scores across all subjects. 

In neuroimaging analyses, rather than binary segmentation masks, total volumes of segmented regions are often used as primary variables of interest. For this reason, we also calculated metrics to evaluate the accuracy of total volume estimations from the segmentations. We used the concordance correlation coefficient ($\rho_c$), a metric that measures the agreement between two variables and that is commonly applied for evaluating reproducibility or inter-rater consistency. We reported the $\rho_c$ between automated and ground-truth segmentation volumes for all subjects.

The metrics that are used in our evaluations are briefly described below.

The $F_1$ score or Dice coefficient \citep{dice1945measures} is a spatial overlap statistic used to gauge the similarity of two segmentations. It's defined as:

\begin{equation*}
  DSC = F_1 = \frac{2 | X \cap Y |}{|X|+|Y|}=\frac{2TP}{2TP+FP+FN}
\end{equation*}

\noindent where $X$ and $Y$ are the predicted and ground truth labels, $TP$, $TN$, $FP$ and $FN$ are the true positive, true negative, false positive and false negative rates respectively.

The $F_2$ score is commonly used in applications where recall is more important than precision (as compared to $F_1$):

\begin{equation*}
  F_2 = \frac{5TP}{5TP+4FN+FP}
\end{equation*}

Considering that our target datasets may typically be imbalanced, i.e. the foreground (target segmentation) may be much smaller compared to the background, we also report the balanced accuracy ($BACC$), which is defined as:

\begin{equation*}
  BACC = \frac{TPR+TNR}{2}
\end{equation*}

\noindent where $TPR=\frac{TP}{TP+FN}$ and $TNR=\frac{TN}{TN+FP}$

The concordance correlation coefficient is defined as:

\begin{equation*}
  \rho_c = \frac{2 \rho \sigma_x \sigma_y}{\sigma_x^2 + \sigma_y^2 + (\mu_x - \mu_y)^2}
\end{equation*}

\noindent where $\mu_x$ and $\mu_y$ are the means and $\sigma_x$ and $\sigma_y$ are the variances of the two variables, and $\rho$ is the correlation correlation between them. 

\section{Experimental Results}

\subsection{White matter lesion segmentation}

\begin{figure}[h]
  \begin{center}
    \includegraphics[width=13cm]{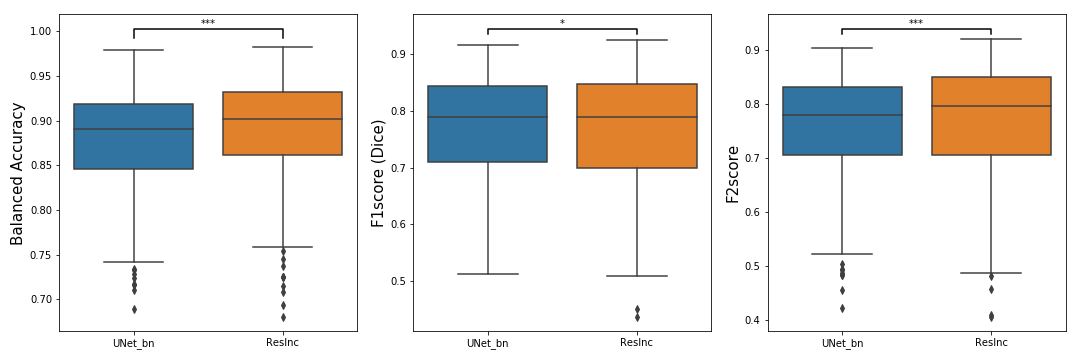}
    \caption{Distribution of scores for the 3 evaluation metrics for segmentation of WML using UNet and DeepMRSeg models}
    \label{Fig_wml}
  \end{center}
\end{figure}

The distribution of balanced accuracy, $F_1$ and $F_2$  scores for the segmentations obtained using UNet and DeepMRSeg are shown in figure \ref{Fig_wml}. DeepMRSeg obtained a significantly better balanced accuracy and $F_2$ score. The mean Dice ($F_1$) score for both methods was similar with no significant differences. DeepMRSeg also obtained a significantly higher $\rho_c$ score (Table \ref{table1}).

\begin{table}[htbp]
\caption{Scores for the 4 evaluation metrics for segmentation of WML using UNet and DeepMRSeg models. For the three overlap metrics, $BACC$, $F_1$ and $F_2$, we report mean and standard deviation across all subjects.}
\resizebox{\textwidth}{!}{
\begin{tabular}{lcccccc|cc}
& \multicolumn{ 2}{c}{$\boldsymbol{BACC}$} & \multicolumn{ 2}{c}{$\boldsymbol{F_1}$} & \multicolumn{ 2}{c}{$\boldsymbol{F_2}$} & \multicolumn{ 2}{c}{$\boldsymbol{\rho_c}$} \\
\toprule
 & \textbf{Unet} & \textbf{DeepMRSeg} & \textbf{Unet} & \textbf{DeepMRSeg} & \textbf{Unet} & \textbf{DeepMRSeg} & \textbf{Unet} & \textbf{DeepMRSeg} \\
\midrule
WML & 0.876 (0.06) & \textbf{0.889 (0.06)}	& \textbf{0.768 (0.10)} & 0.765 (0.10)	& 0.759 (0.10) & \textbf{0.769 (0.10)}	&  0.956 &  \textbf{0.962} \\
\bottomrule
\end{tabular}
}
\label{table1}
\end{table}

\subsection{Mid-brain segmentation}
\begin{figure}[h]
  \begin{center}
    \includegraphics[width=13cm]{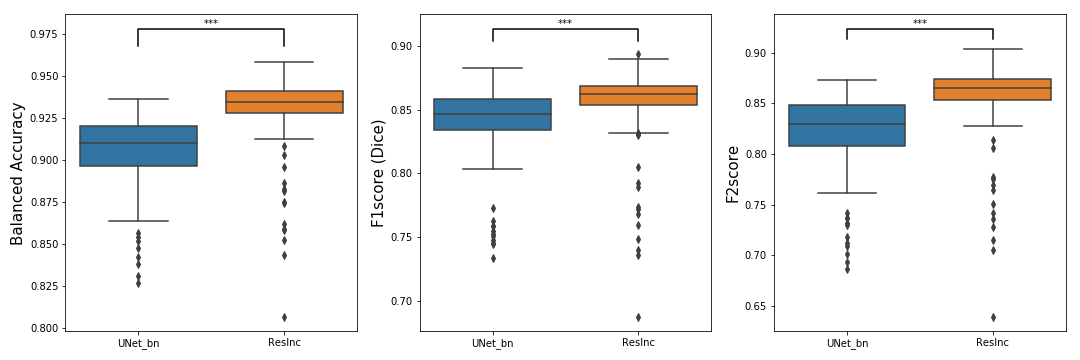}
    \caption{Distribution of scores for the 3 evaluation metrics for segmentation of deep brain structures using UNet and DeepMRSeg models}
    \label{Fig_midbrain}
  \end{center}
\end{figure}

UNet and DeepMRSeg networks were applied for segmenting each scan into 14 target ROIs. We calculated evaluation scores for each ROI independently. The distribution of overlap scores over all ROIs and subjects are shown in  figure \ref{Fig_midbrain}. 
DeepMRSeg obtained a significantly higher balanced accuracy for each ROI indpendently, as well as on average across all ROIs. DeepMRSeg also obtained higher $\rho_c$ for all ROIs except left caudate and left and right pallidum (table \ref{table2}).

\begin{table}[htbp]
\caption{Scores for the 4 evaluation metrics for segmentation of deep brain structures using UNet and DeepMRSeg models. For the three overlap metrics, $BACC$, $F_1$ and $F_2$, we report mean and standard deviation across all subjects.}
\resizebox{\textwidth}{!}{
\begin{tabular}{lcccccc|cc}
& \multicolumn{ 2}{c}{$\boldsymbol{BACC}$} & \multicolumn{ 2}{c}{$\boldsymbol{F_1}$} & \multicolumn{ 2}{c}{$\boldsymbol{F_2}$} & \multicolumn{ 2}{c}{$\boldsymbol{\rho_c}$} \\
\toprule
\textbf{ROI} & \textbf{Unet} & \textbf{DeepMRSeg} & \textbf{Unet} & \textbf{DeepMRSeg} & \textbf{Unet} & \textbf{DeepMRSeg} & \textbf{Unet} & \textbf{DeepMRSeg} \\
\midrule
Right Accumbens Area & 0.867 (0.05) & \textbf{0.894 (0.05)} & 0.762 (0.05) & \textbf{0.765 (0.07)} & 0.743 (0.07) & \textbf{0.777 (0.09)} & 0.682  & \textbf{0.764} \\ 
Left Accumbens Area & 0.855 (0.05) & \textbf{0.896 (0.05)} & 0.755 (0.06) & \textbf{0.775 (0.06)} & 0.725 (0.08) & \textbf{0.784 (0.08)} & 0.565  & \textbf{0.808} \\ 
Right Amygdala & 0.845 (0.05) & \textbf{0.88 (0.03)} & 0.751 (0.06) & \textbf{0.782 (0.04)} & 0.712 (0.08) & \textbf{0.767 (0.05)} & 0.266  & \textbf{0.466} \\ 
Left Amygdala & 0.855 (0.04) & \textbf{0.889 (0.03)} & 0.764 (0.05) & \textbf{0.798 (0.04)} & 0.73 (0.07) & \textbf{0.785 (0.05)} & 0.349  & \textbf{0.592} \\ 
Right Caudate & 0.928 (0.04) & \textbf{0.945 (0.05)} & 0.883 (0.06) & \textbf{0.893 (0.07)} & 0.866 (0.08) & \textbf{0.89 (0.08)} & 0.802  & \textbf{0.854} \\ 
Left Caudate & 0.943 (0.02) & \textbf{0.948 (0.03)} & 0.891 (0.04) & \textbf{0.897 (0.05)} & 0.887 (0.04) & \textbf{0.896 (0.05)} & \textbf{0.912 } & 0.900 \\ 
Right Hippocampus & 0.892 (0.03) & \textbf{0.924 (0.03)} & 0.833 (0.03) & \textbf{0.858 (0.03)} & 0.802 (0.05) & \textbf{0.852 (0.04)} & 0.425  & \textbf{0.736} \\ 
Left Hippocampus & 0.893 (0.03) & \textbf{0.921 (0.03)} & 0.832 (0.03) & \textbf{0.856 (0.03)} & 0.803 (0.05) & \textbf{0.847 (0.04)} & 0.396  & \textbf{0.662} \\ 
Right Pallidum & 0.916 (0.04) & \textbf{0.952 (0.03)} & \textbf{0.854 (0.05)} & 0.859 (0.04) & 0.84 (0.06) & \textbf{0.886 (0.04)} & \textbf{0.778 } & 0.687 \\ 
Left Pallidum & 0.927 (0.05) & \textbf{0.955 (0.03)} & 0.857 (0.06) & \textbf{0.856 (0.04)} & 0.854 (0.08) & \textbf{0.886 (0.04)} & \textbf{0.674 } & 0.473 \\ 
Right Putamen & 0.941 (0.02) & \textbf{0.953 (0.02)} & 0.901 (0.03) & \textbf{0.908 (0.03)} & 0.889 (0.04) & \textbf{0.906 (0.04)} & 0.863  & \textbf{0.899} \\ 
Left Putamen & 0.939 (0.03) & \textbf{0.955 (0.03)} & 0.899 (0.04) & \textbf{0.907 (0.04)} & 0.886 (0.05) & \textbf{0.908 (0.05)} & 0.815  & \textbf{0.894} \\ 
Right Thalamus Proper & 0.936 (0.02) & \textbf{0.959 (0.02)} & 0.9 (0.02) & \textbf{0.914 (0.01)} & 0.883 (0.04) & \textbf{0.916 (0.02)} & 0.752 & \textbf{0.906} \\ 
Left Thalamus Proper & 0.946 (0.02) & \textbf{0.954 (0.02)} & 0.906 (0.01) & \textbf{0.912 (0.01)} & 0.898 (0.03) & \textbf{0.91 (0.02)} & 0.849 & \textbf{0.888} \\ 
\midrule
\textbf{Average} & 0.906 (0.02) & \textbf{0.93 (0.02)} & 0.842 (0.03) & \textbf{0.856 (0.03)} & 0.823 (0.04) & \textbf{0.858 (0.04)} & 0.987 & \textbf{0.993} \\
\bottomrule
\end{tabular}
}
\label{table2}
\end{table}

\subsection{Hippocampus segmentation}
\begin{figure}[h]
  \begin{center}
    \includegraphics[width=13cm]{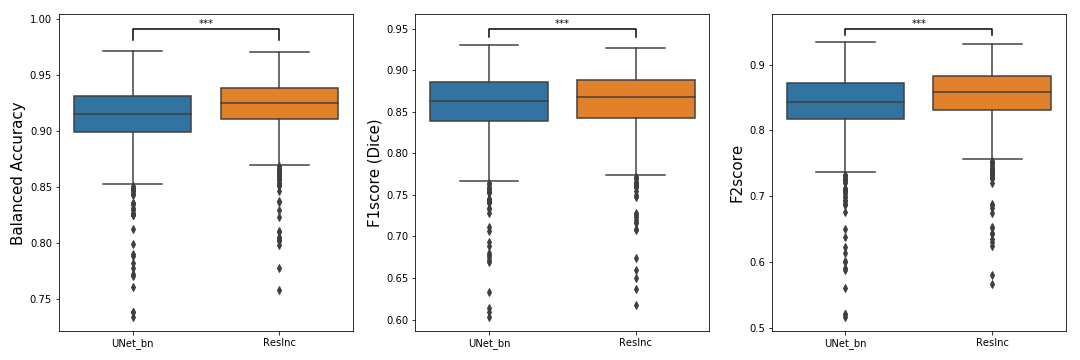}
    \caption{Distribution of scores for the 3 evaluation metrics for segmentation of hippocampus sub-regions using UNet and DeepMRSeg models}
    \label{Fig_hippo}
  \end{center}
\end{figure}

The hippocampus was segmented into two sub-regions using UNet and DeepMRSeg. We calculated overlap scores for each sub-region independently. DeepMRSeg obtained a significantly better accuracy for both hippocampus sub-regions (Figure \ref{Fig_hippo} and table \ref{table3}).

\begin{table}[htbp]
\caption{Scores for the 4 evaluation metrics for segmentation of hippocampus sub-regions using UNet and DeepMRSeg models. For the three overlap metrics, $BACC$, $F_1$ and $F_2$, we report mean and standard deviation across all subjects.}
\resizebox{\textwidth}{!}{
\begin{tabular}{lcccccc|cc}
& \multicolumn{ 2}{c}{$\boldsymbol{BACC}$} & \multicolumn{ 2}{c}{$\boldsymbol{F_1}$} & \multicolumn{ 2}{c}{$\boldsymbol{F_2}$} & \multicolumn{ 2}{c}{$\boldsymbol{\rho_c}$} \\
\toprule
\textbf{ROI} & \textbf{Unet} & \textbf{DeepMRSeg} & \textbf{Unet} & \textbf{DeepMRSeg} & \textbf{Unet} & \textbf{DeepMRSeg} & \textbf{Unet} & \textbf{DeepMRSeg} \\
\midrule
Anterior &  0.917 (0.03) & \textbf{ 0.925 (0.03)} &  0.866 (0.04) & \textbf{ 0.869 (0.04)} &  0.848 (0.06) & \textbf{ 0.858 (0.05)} &  0.765  & \textbf{ 0.800 } \\
Posterior &  0.908 (0.03) & \textbf{ 0.920 (0.03)} &  0.849 (0.05) & \textbf{ 0.858 (0.04)} &  0.830 (0.06) & \textbf{ 0.847 (0.05)} &  0.624 & \textbf{ 0.734 } \\
\midrule
\textbf{Average} & 0.913 (0.03) & \textbf{0.922 (0.02)} & 0.857 (0.04) & \textbf{0.862 (0.04)} & 0.839 (0.05) & \textbf{0.852 (0.04)} & 0.726  & \textbf{0.786}  \\
\bottomrule
\end{tabular}
}
\label{table3}
\end{table}

\section{Conclusions}
We presented a novel deep learning based MRI segmentation method that combines elements from recent major advances in the field. The proposed network architecture was built with two main motivations: providing a generic tool that can be used for different types of segmentation problems, rather than being specific to a single type of target label; and allowing a wide range of users to easily segment their images by directly using their raw T1 scans without need for any pre-processing steps. In our validation experiments, we showed that DeepMRSeg outperformed a standard UNet implementation used as benchmark in three different segmentation tasks, achieving highly accurate segmentations in all tasks. We provide code and pre-trained models that can be used for applying DeepMRSeg on new scans.

\bibliographystyle{elsarticle-harv}
\bibliography{biblio_DeepMRSeg_vArxiv.2}

\begin{thebibliography}{25}
\expandafter\ifx\csname natexlab\endcsname\relax\def\natexlab#1{#1}\fi
\expandafter\ifx\csname url\endcsname\relax
  \def\url#1{\texttt{#1}}\fi
\expandafter\ifx\csname urlprefix\endcsname\relax\def\urlprefix{URL }\fi

\bibitem[{Akkus et~al.(2017)Akkus, Galimzianova, Hoogi, Rubin, and
  Erickson}]{Akkus17}
Akkus, Z., Galimzianova, A., Hoogi, A., Rubin, D.~L., Erickson, B.~J., Aug
  2017. Deep learning for brain mri segmentation: State of the art and future
  directions. Journal of Digital Imaging 30~(4), 449--459.

\bibitem[{Anbeek et~al.(2004)Anbeek, Vincken, van Osch, Bisschops, and van~der
  Grond}]{ANBEEK20041037}
Anbeek, P., Vincken, K.~L., van Osch, M.~J., Bisschops, R.~H., van~der Grond,
  J., 2004. Probabilistic segmentation of white matter lesions in mr imaging.
  NeuroImage 21~(3), 1037 -- 1044.

\bibitem[{Anwar et~al.(2018)Anwar, Majid, Qayyum, Awais, Alnowami, and
  Khan}]{Anwar18}
Anwar, S.~M., Majid, M., Qayyum, A., Awais, M., Alnowami, M., Khan, M.~K., Nov.
  2018. Medical image analysis using convolutional neural networks: A review.
  J. Med. Syst. 42~(11), 1--13.
\newline\urlprefix\url{https://doi.org/10.1007/s10916-018-1088-1}

\bibitem[{Asman et~al.(2013)Asman, Akhondi-Asl, Wang, Tustison, Avants,
  Warfield, and Landman}]{asman2013miccai}
Asman, A., Akhondi-Asl, A., Wang, H., Tustison, N., Avants, B., Warfield,
  S.~K., Landman, B., 2013. Miccai 2013 segmentation algorithms, theory and
  applications (sata) challenge results summary. In: MICCAI Challenge Workshop
  on Segmentation: Algorithms, Theory and Applications (SATA).

\bibitem[{Despotovic et~al.(2015)Despotovic, Goossens, and
  Philips}]{despotovic15}
Despotovic, I., Goossens, B., Philips, W., 2015. {MRI} segmentation of the
  human brain: Challenges, methods, and applications. Comp. Math. Methods in
  Medicine 2015, 450341:1--450341:23.

\bibitem[{Dice(1945)}]{dice1945measures}
Dice, L.~R., 1945. Measures of the amount of ecologic association between
  species. Ecology 26~(3), 297--302.

\bibitem[{Eugenio~Iglesias and Sabuncu(2014)}]{iglesias14}
Eugenio~Iglesias, J., Sabuncu, M., 12 2014. Multi-atlas segmentation of
  biomedical images: A survey. Medical image analysis 24.

\bibitem[{Gonzalez-Villa et~al.(2016)Gonzalez-Villa, Oliver, Valverde, Wang,
  Zwiggelaar, and Llado}]{GONZALEZVILLA201645}
Gonzalez-Villa, S., Oliver, A., Valverde, S., Wang, L., Zwiggelaar, R., Llado,
  X., 2016. A review on brain structures segmentation in magnetic resonance
  imaging. Artificial Intelligence in Medicine 73, 45 -- 69.

\bibitem[{Gordillo et~al.(2013)Gordillo, Montseny, and
  Sobrevilla}]{GORDILLO20131426}
Gordillo, N., Montseny, E., Sobrevilla, P., 2013. State of the art survey on
  mri brain tumor segmentation. Magnetic Resonance Imaging 31~(8), 1426 --
  1438.

\bibitem[{Guo et~al.(2018)Guo, Liu, Georgiou, and Lew}]{Guo2018}
Guo, Y., Liu, Y., Georgiou, T., Lew, M.~S., Jun 2018. A review of semantic
  segmentation using deep neural networks. International Journal of Multimedia
  Information Retrieval 7~(2), 87--93.

\bibitem[{He et~al.(2015)He, Zhang, Ren, and Sun}]{HeZRS15}
He, K., Zhang, X., Ren, S., Sun, J., 2015. Deep residual learning for image
  recognition. arXiv preprint abs/1512.03385.

\bibitem[{Ioffe and Szegedy(2015)}]{IoffeS15}
Ioffe, S., Szegedy, C., 2015. Batch normalization: Accelerating deep network
  training by reducing internal covariate shift. arXiv preprint abs/1502.03167.
\newline\urlprefix\url{http://arxiv.org/abs/1502.03167}

\bibitem[{Kalavathi~P(2016)}]{Kalavathi16}
Kalavathi~P, P.~V., 2016. Methods on skull stripping of mri head scan images-a
  review. J Digit Imaging 29~(3), 365 -- 379.

\bibitem[{Kamnitsas et~al.(2016)Kamnitsas, Ledig, Newcombe, Simpson, Kane,
  Menon, Rueckert, and Glocker}]{KamnitsasLNSKMR16}
Kamnitsas, K., Ledig, C., Newcombe, V. F.~J., Simpson, J.~P., Kane, A.~D.,
  Menon, D.~K., Rueckert, D., Glocker, B., 2016. Efficient multi-scale 3d {CNN}
  with fully connected {CRF} for accurate brain lesion segmentation. arXiv
  preprint abs/1603.05959.
\newline\urlprefix\url{http://arxiv.org/abs/1603.05959}

\bibitem[{Krizhevsky et~al.(2012)Krizhevsky, Sutskever, and
  Hinton}]{NIPS2012_4824}
Krizhevsky, A., Sutskever, I., Hinton, G.~E., 2012. Imagenet classification
  with deep convolutional neural networks. In: Pereira, F., Burges, C. J.~C.,
  Bottou, L., Weinberger, K.~Q. (Eds.), Advances in Neural Information
  Processing Systems 25. Curran Associates, Inc., pp. 1097--1105.

\bibitem[{{Kuijf} et~al.(2019){Kuijf}, {Biesbroek}, {de Bresser}, {Heinen},
  {Andermatt}, {Bento}, {Berseth}, {Belyaev}, {Cardoso}, {Casamitjana},
  {Collins}, {Dadar}, {Georgiou}, {Ghafoorian}, {Jin}, {Khademi}, {Knight},
  {Li}, {Lladó}, {Luna}, {Mahmood}, {McKinley}, {Mehrtash}, {Ourselin},
  {Park}, {Park}, {Park}, {Pezold}, {Puybareau}, {Rittner}, {Sudre},
  {Valverde}, {Vilaplana}, {Wiest}, {Xu}, {Xu}, {Zeng}, {Zhang}, {Zheng},
  {Chen}, {van der Flier}, {Barkhof}, {Viergever}, and {Biessels}}]{Kuijf19}
{Kuijf}, H.~J., {Biesbroek}, J.~M., {de Bresser}, J., {Heinen}, R.,
  {Andermatt}, S., {Bento}, M., {Berseth}, M., {Belyaev}, M., {Cardoso}, M.~J.,
  {Casamitjana}, A., {Collins}, D.~L., {Dadar}, M., {Georgiou}, A.,
  {Ghafoorian}, M., {Jin}, D., {Khademi}, A., {Knight}, J., {Li}, H., {Lladó},
  X., {Luna}, M., {Mahmood}, Q., {McKinley}, R., {Mehrtash}, A., {Ourselin},
  S., {Park}, B., {Park}, H., {Park}, S.~H., {Pezold}, S., {Puybareau}, E.,
  {Rittner}, L., {Sudre}, C.~H., {Valverde}, S., {Vilaplana}, V., {Wiest}, R.,
  {Xu}, Y., {Xu}, Z., {Zeng}, G., {Zhang}, J., {Zheng}, G., {Chen}, C., {van
  der Flier}, W., {Barkhof}, F., {Viergever}, M.~A., {Biessels}, G.~J., 2019.
  Standardized assessment of automatic segmentation of white matter
  hyperintensities; results of the wmh segmentation challenge. IEEE
  Transactions on Medical Imaging, 1--1.

\bibitem[{Mu and Gage(2011)}]{Mu2011AdultHN}
Mu, Y., Gage, F.~H., 2011. Adult hippocampal neurogenesis and its role in
  alzheimer's disease. In: Molecular Neurodegeneration.

\bibitem[{Nair and Hinton(2010)}]{Nair10}
Nair, V., Hinton, G.~E., 2010. Rectified linear units improve restricted
  boltzmann machines. In: Proceedings of the 27th International Conference on
  International Conference on Machine Learning. ICML'10. Omnipress, USA, pp.
  807--814.

\bibitem[{Ronneberger et~al.(2015)Ronneberger, Fischer, and
  Brox}]{RonnebergerFB15}
Ronneberger, O., Fischer, P., Brox, T., 2015. U-net: Convolutional networks for
  biomedical image segmentation. arXiv preprint arXiv:1505.04597.

\bibitem[{Simpson et~al.(2019)Simpson, Antonelli, Bakas, Bilello, Farahani, van
  Ginneken, Kopp{-}Schneider, Landman, Litjens, Menze, Ronneberger, Summers,
  Bilic, Christ, Do, Gollub, Golia{-}Pernicka, Heckers, Jarnagin, McHugo,
  Napel, Vorontsov, Maier{-}Hein, and Cardoso}]{simpson19}
Simpson, A.~L., Antonelli, M., Bakas, S., Bilello, M., Farahani, K., van
  Ginneken, B., Kopp{-}Schneider, A., Landman, B.~A., Litjens, G. J.~S., Menze,
  B.~H., Ronneberger, O., Summers, R.~M., Bilic, P., Christ, P.~F., Do, R.
  K.~G., Gollub, M., Golia{-}Pernicka, J., Heckers, S., Jarnagin, W.~R.,
  McHugo, M., Napel, S., Vorontsov, E., Maier{-}Hein, L., Cardoso, M.~J., 2019.
  A large annotated medical image dataset for the development and evaluation of
  segmentation algorithms. arXiv preprint abs/1902.09063.
\newline\urlprefix\url{http://arxiv.org/abs/1902.09063}

\bibitem[{Springenberg et~al.(2014)Springenberg, Dosovitskiy, Brox, and
  Riedmiller}]{springenberg2014striving}
Springenberg, J.~T., Dosovitskiy, A., Brox, T., Riedmiller, M., 2014. Striving
  for simplicity: The all convolutional net. arXiv preprint arXiv:1412.6806.

\bibitem[{Szegedy et~al.(2016)Szegedy, Ioffe, and Vanhoucke}]{SzegedyIV16}
Szegedy, C., Ioffe, S., Vanhoucke, V., 2016. Inception-v4, inception-resnet and
  the impact of residual connections on learning. arXiv preprint
  abs/1602.07261.
\newline\urlprefix\url{http://arxiv.org/abs/1602.07261}

\bibitem[{Szegedy et~al.(2014)Szegedy, Liu, Jia, Sermanet, Reed, Anguelov,
  Erhan, Vanhoucke, and Rabinovich}]{SzegedyLJSRAEVR14}
Szegedy, C., Liu, W., Jia, Y., Sermanet, P., Reed, S.~E., Anguelov, D., Erhan,
  D., Vanhoucke, V., Rabinovich, A., 2014. Going deeper with convolutions.
  arXiv preprint abs/1409.4842.
\newline\urlprefix\url{http://arxiv.org/abs/1409.4842}

\bibitem[{Szegedy et~al.(2015)Szegedy, Vanhoucke, Ioffe, Shlens, and
  Wojna}]{SzegedyVISW15}
Szegedy, C., Vanhoucke, V., Ioffe, S., Shlens, J., Wojna, Z., 2015. Rethinking
  the inception architecture for computer vision. arXiv preprint
  abs/1512.00567.
\newline\urlprefix\url{http://arxiv.org/abs/1512.00567}

\bibitem[{Wardlaw et~al.(2013)Wardlaw, Smith, J~Biessels, Cordonnier, Fazekas,
  Frayne, Lindley, O'Brien, Barkhof, R~Benavente, E~Black, Brayne, Breteler,
  Chabriat, DeCarli, De~Leeuw, Doubal, Duering, C~Fox, and for ReportIng
  Vascular changes on~nEuroimaging (STRIVE)}]{Wardlaw13}
Wardlaw, J., Smith, E., J~Biessels, G., Cordonnier, C., Fazekas, F., Frayne,
  R., Lindley, R., O'Brien, J., Barkhof, F., R~Benavente, O., E~Black, S.,
  Brayne, C., Breteler, M., Chabriat, H., DeCarli, C., De~Leeuw, F.-E., Doubal,
  F., Duering, M., C~Fox, N., for ReportIng Vascular changes on~nEuroimaging
  (STRIVE), 2013. Neuroimaging standards for research into small vessel disease
  and its contribution to ageing and neurodegeneration. The Lancet Neurology
  12, 822--838.

\end{thebibliography}

\end{document}